\documentclass[twocolumn]{aa}
\usepackage{txfonts}
\usepackage{graphicx}

\begin{document}

   \title{Electron density and carriers of the diffuse interstellar bands}

   \subtitle{}

   \author{P. Gnaci\'nski \inst{1},
          J.K. Sikorski \inst{2},
          G.A. Galazutdinov \inst{3}
          }
   \authorrunning{P. Gnaci\'nski {\it et al.}}
          
   \offprints{P. Gnaci\'nski}

   \institute{Institute of Theoretical Physics and Astrophysics,
              University of Gda\'nsk,
              ul. Wita Stwosza 57, 80-952 Gda\'nsk, Poland\\
              \email{pg@iftia.univ.gda.pl}
         \and
            Institute of Experimental Physics,
            University of Gda\'nsk,
            ul. Wita Stwosza 57, 80-952 Gda\'nsk, Poland\\
            \email{fizjks@iftia.univ.gda.pl}
         \and
            Korea Astronomy and Space Science Institute,
            61-1 Whaam-dong, Yuseong-gu, 
            Daejeon, 305-348, Republic of Korea \\
            \email{gala@boao.re.kr}
         }

   \date{}

   \abstract{
  
     We have used the ionisation equilibrium equation to derive the electron density in interstellar clouds in the direction to 13 stars. A linear relation was found, that allows the determination of the electron density from the Mg I and Mg II column densities in  diffuse clouds.
   The comparison of normalised equivalent width of 12 DIBs with the electron density shows that the DIBs equivalent width do not change with electron density varying 
in the range $n_e=0.01\div 2.5$ cm$^{-3}$. 
Therefore the DIBs carriers (1) can be observed only in one ionisation stage, or (2)
the DIBs are arising in cloud regions (eg. cores or cloud coronas) for which we can not determine the electron density.   
   
   }

   \maketitle

\section{Introduction}

  The Diffuse Interstellar Bands (DIBs) are broad absorption features, seen in the interstellar medium. There are almost 300 DIBs known in the optical and NIR spectrum (\cite{Gazinur}). Despite of over 80 years of investigations their nature is still unknown (for a review, see \cite{Herbig}). Many carriers have been proposed as the carriers of DIBs, e.g. solid particles, simple molecules, negative atomic ions,  carbon chains, fullerens. 
  
  In 1985 \cite{Zwet} proposed polycyclic aromatic hydrocarbons (PAHs) as the source of DIBs. Since then it is a popular hypothesis. There are however problems with obtaining gas phase laboratory spectra of dehydrogenated and/or ionised PAHs to verify this hypothesis. 
  Recently \cite{Cox} have presented simulations of PAH charge state distribution in various environments. In clouds with various irradiation and density the fractional abundances of PAH cations, neutrals and anions changes dramatically.
  
  We have used the ionisation equilibrium equation to obtain the electron densities in
individual clouds. The electron density have been compared to the equivalent width of 
DIBs and the CH/CH+ lines. The equivalent width of the CH+ line drops with rising $n_e$,
but no changes of the DIBs equivalent widths are observed.

  
\section{Column densities and equivalent widths}

   The aim of this paper is to check the dependiences between the electron density and the carriers of DIBs. In order to determine the electron density we had to measure the column densities of two elements in two adjacent ionisation stages.
   Our target stars were stars fulfilling the following criteria:
   
\begin{itemize}
  \item reddened stars of spectral type O or B
  \item high resolution Hubble Space Telescope (HST) spectra for at least MgI, MgII lines are available (both, GHRS and STIS spectra were used)
  \item hydrogen column densities are available
  \item equivalent widths of the chosen DIBs are available
\end{itemize}
   Column densities of Mg I, Mg II, Si I, Si II, C II, C II* were calculated from high resolution HST spectra.
    The spectra from the ultraviolet spectral range were downloaded from the HST Data Archive. The GHRS spectra taken in the FP-SPLIT mode were processed with IRAF tasks {\it poffsets} and {\it specalign} to achieve the final spectrum. The column densities were derived using the profile fitting technique. The absorption lines were fitted by Voigt profiles. The transitions for which the natural dumping constant ($\Gamma$) in not known (Mg II 1240 \AA\ doublet, Mg I 1828 \AA ) were fitted with a Gauss function. The cloud velocities (v), Doppler broadening parameters (b) and column densities (N) for multiple absorption components were simultaneously fitted to the observed spectrum. Both lines of magnesium doublet (at 1200 \AA) were also fitted simultaneously - v, b and N were common for both lines in the doublet. The wavelengths, oscillator strengths (f) and natural damping constants ($\Gamma$) were adopted from \cite{Morton}. 
    
    A convolution with a point spread function (PSF) was also performed. The PSF for the GHRS spectrograph consists of two Gaussian components. The "core" Gaussian has a FWHM=1.05 diodes, while the "halo" component has FWHM=5.0 diodes (\cite{Spitzer}). 
The relative contribution of the "core" and "halo" components into the PSF is wavelength dependent and was interpolated from the table given by \cite{Cardelli}. 
The Gaussian PSF for the STIS spectrograph depends on wavelength, slit and the mode of observations. The tables with FWHM for the combination of mode and slit can be found in "STIS Instrument Handbook" (\cite{Kim}).
The FWHM of the Gaussian PSF was wavelength-interpolated from these tables. 
    
    The derived column densities, used to calculate the electron density are presented in Table 1. The hydrogen (HI) column densities were adopted from \cite{Dip}. Molecular hydrogen (H$_2$) column densities come from \cite{Rach} and \cite{Sav}. The equivalent widths of  DIBs and CH/CH+ were kindly supported by Jacek Kre\l owski.

\section{Electron density}

  The electron density ($n_e$ in $cm^{-3}$) was calculated from the equations of ionisation equilibrium for two elements. The first element was Mg, because it is easily observed in two ionisation stages. The Mg II column density was determined from the 1240 \AA\ doublet, the Mg I column density was determined from the 2026 \AA, 2852 \AA\ or 1827 \AA\ line. 
  
\begin{figure*}
   \centering
   \includegraphics[width=\textwidth]{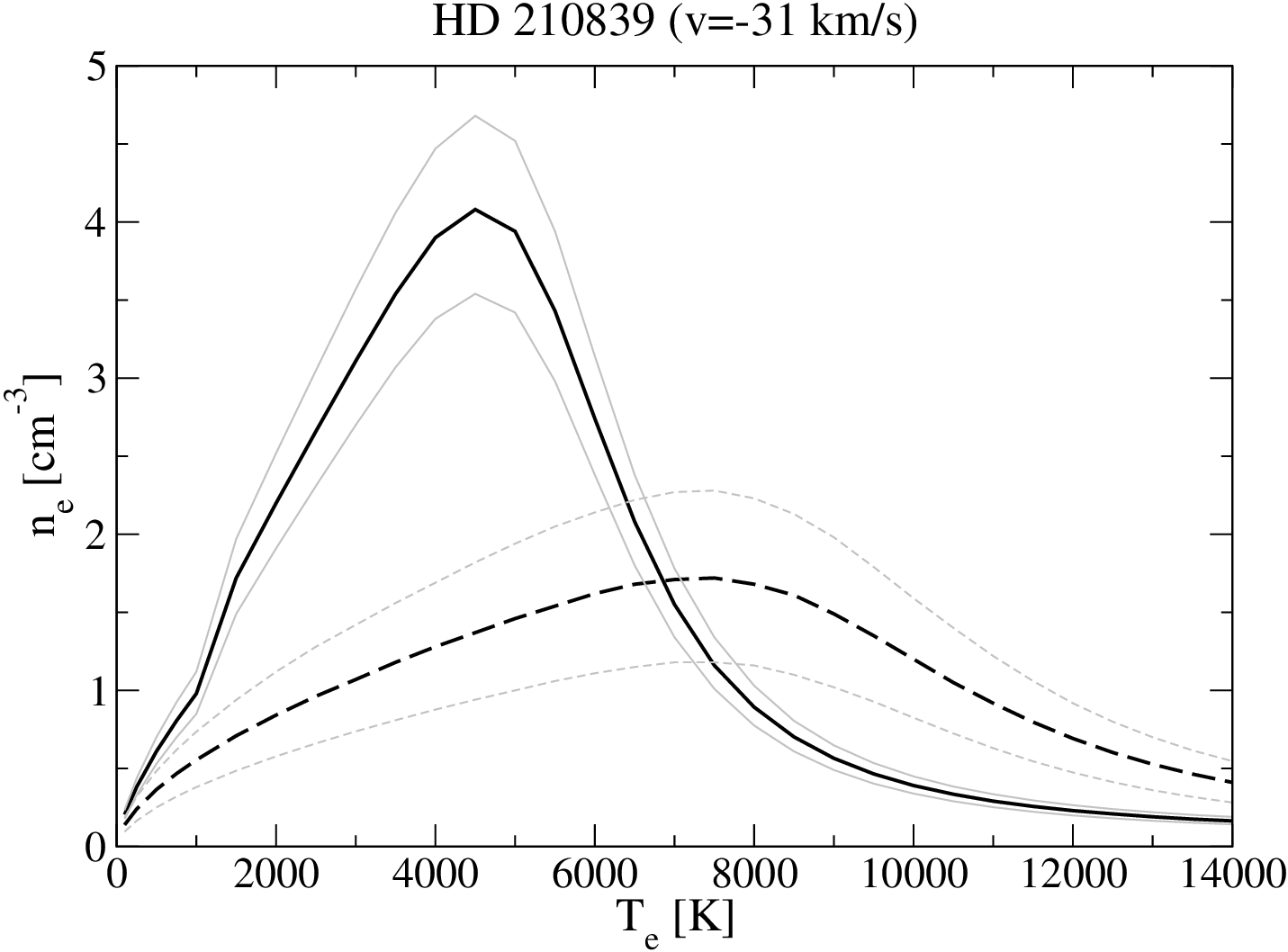}
      \caption{  The $n_{e}(T_{e})$ curves from MgI/MgII (solid line) and SiI/SiII (dashed                    line) for the $v=-31\ km/s$ cloud in the direction of HD 210839.
                 The intersection point of this curves allows us to determine the 
                 electron density ($n_{e}$).
              }
      \label{Intersection}
      \vspace{1cm}
   \end{figure*}

  The step rise of dielectron recombination coefficient for Mg II with temperature causes the decrease of electron density, inferred from MgI/MgII, with temperature (Fig. \ref{Intersection}). Such behaviour enables calculation of electron density, because the curve $n_e(T_{e})$ from MgI/MgII intersects with a curve $n_e(T_{e})$ from another element. The equation of ionisation equilibrium for Mg is the following:
  \begin{equation}
    \frac{n_{e}N(Mg\: II)}{N(Mg\: I)}=\frac{\Gamma(Mg_{12})+n_{e}C(Mg_{12})}{\alpha_{rad}(Mg_{21})+\alpha_{die}(Mg_{21})}
  \end{equation}
Where $N(MgII)$ and $N(MgI)\ [cm^{-2}]$ are the column densities of ionised and neutral Mg; $\alpha_{rad}(Mg_{21})\ [cm^3/s]$ is the radiative recombination coefficient; $\alpha_{die}(Mg_{21})\ [cm^3/s]$ is the dielectronic recombination rate; 
$\Gamma(Mg_{12})\ [1/s]$ is the ionisation rate of MgI by UV photons; $C(Mg_{12})\ [cm^3/s]$ is the collisional ionisation rate.

  Because the coefficients $\alpha_{rad}$, $\alpha_{die}$ and $C$ depend from electron temperature ($T_e$) we need an analogous equation for a second element to obtain $n_e$ and $T_e$ simultaneously. For stars: HD 24534, HD 203374, HD 206267, HD 209339, HD 210839, the second element was Si. The column density of Si I was calculated form the 1845 \AA \  line, and column density of Si II from the 1808  \AA \ one. From the intersection of the $n_e(T)$ curves from Mg and Si (Fig. \ref{Intersection}) we have obtained the electron density.
  
  The $\Gamma$ coefficients for Mg and Si were adopted from the WJ2 model (\cite{Boer}). The recombination coefficients ($\alpha_{rad}$ and $\alpha_{die}$) and the collisional ionisation rate coefficient ($C$) (see Table 2) were adopted from \cite{Shull}.
  
  For the stars HD 202904 and HD 160578 the ionisation equilibrium was calculated from MgI/MgII and CII/CII*. The column density of carbon was calculated from the C II 1335 \AA \ and C II* 1336 \AA \ lines, using the profile fitting technique (see \cite{Gna}  for details). The equilibrium between the collisional excitation and radiative de-excitation of ionised carbon is described by the equation (\cite{Wood}): 
  \begin{equation}
    \frac{N(C\: II^*)}{N(C\: II)}=\frac{n_{e}C(C_{12})}{\alpha_{rad}(C_{21})}
  \end{equation}
  The radiative de-excitation $\alpha_{rad}(C_{21})$ was adopted from \cite{NS}. The collision rate coefficient $C(C_{12})$ was adopted from \cite{Wood} and \cite{Hayes}.

  We have also tryed to use Ca as the second element for obtaining $n_e$. Unfortuatelly,
the CaI/CaII ionisation equilibrium curve (eq. 1) for the stars HD 74455 and HD 149757 does not intersect the ionisation equilibrium curve for Mg I/II. The problem can be caused by the change of $n(CaII)/n(CaI)$ between the edge and the centre of the cloud. \cite{Lepp} have found in their numerical simulations, that $n(CaII)/n(CaI)$ decreases from 4800 at the edge to 160 at the centre of the $\zeta$ Persei cloud. 
  
\section{Results and Discussion}
  
\begin{figure*}
   \centering
   \includegraphics[width=\textwidth]{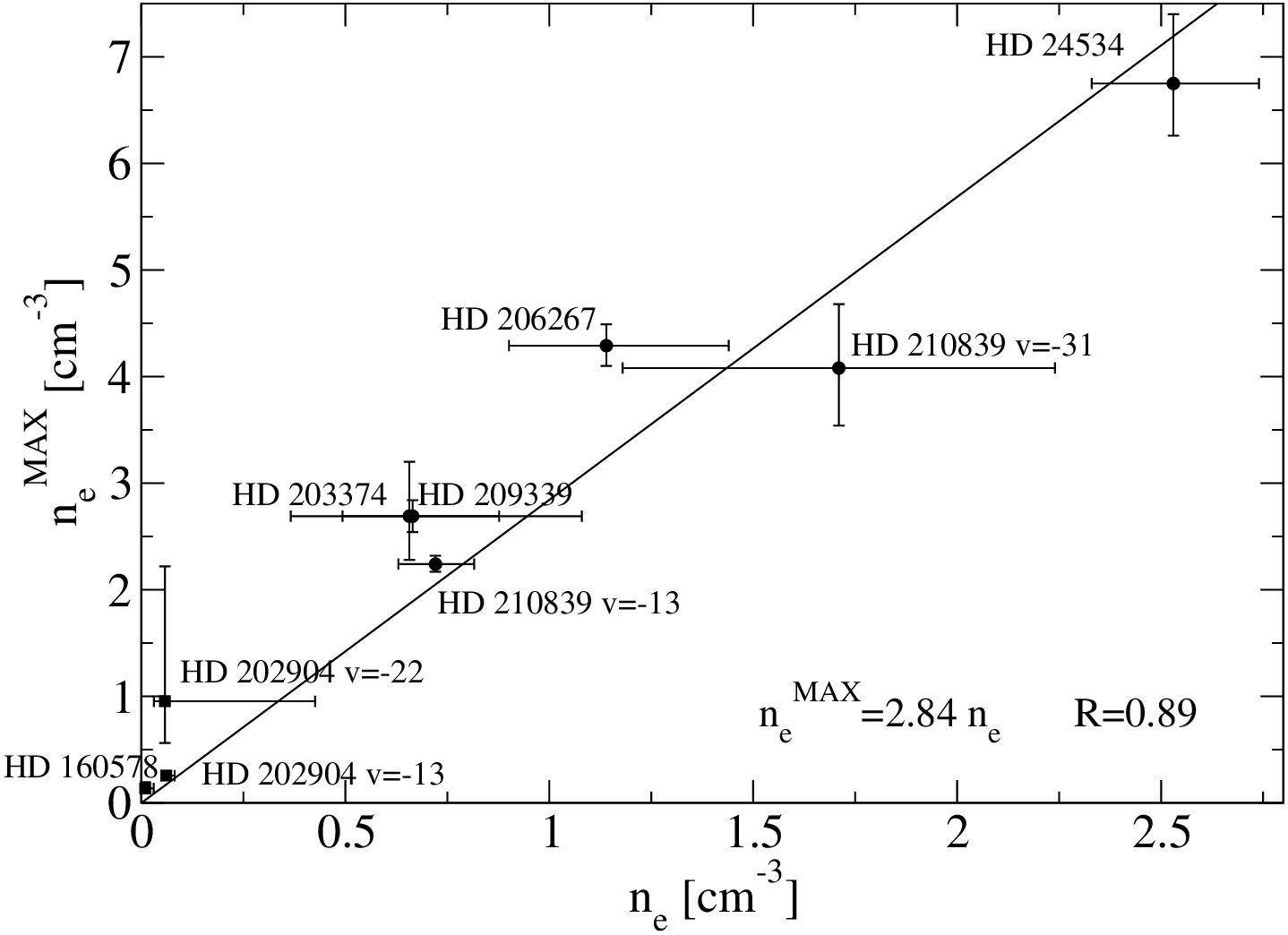}
      \caption{ The linear correlation between maximum $n_e$ from the $N(MgII)/N(MgI)$ 
                curve and the exact $n_e$ value. }
      \label{FigNeMAX}
      \vspace{1cm}
   \end{figure*}

  For a lot of stars Mg is the only element with observations of absorption lines
of two stages of ionisation. From the column densities of neutral and ionised Mg we can only
calculate the maximal possible electron density $n_{e}^{MAX}$. It is simply the maximum of the $n_{e}(T_{e})$ curve. Fortunately this maximum ($n_{e}^{MAX}$) is very well correlated with the electron density $n_{e}$ (Fig. \ref{FigNeMAX}). The correlation coefficient R=0.89, and the linear correlation is following: $n_{e}^{MAX}=2.84 \cdot n_{e}$. 
This relation was derived from $n_e=0.01\div 2.5$ cm$^{-3}$ and may not hold for
denser or thiner environments. The linear relation $n_{e}^{MAX}=2.84 \cdot n_{e}$ 
probably reflects the fact, that most of the clouds for which we can calculate $n_e$ have the electron temperature $T_e\sim 7500 K$. For stars HD 24912, HD 74455, HD 91316, HD 141637, HD 147165, HD 147933, HD 149757 the electron density was calculated using the $n_{e}^{MAX}$ value and this formula. All derived electron densities are presented in Table \ref{tabela}.
  
\begin{figure*}
   \centering
   \includegraphics[width=\textwidth]{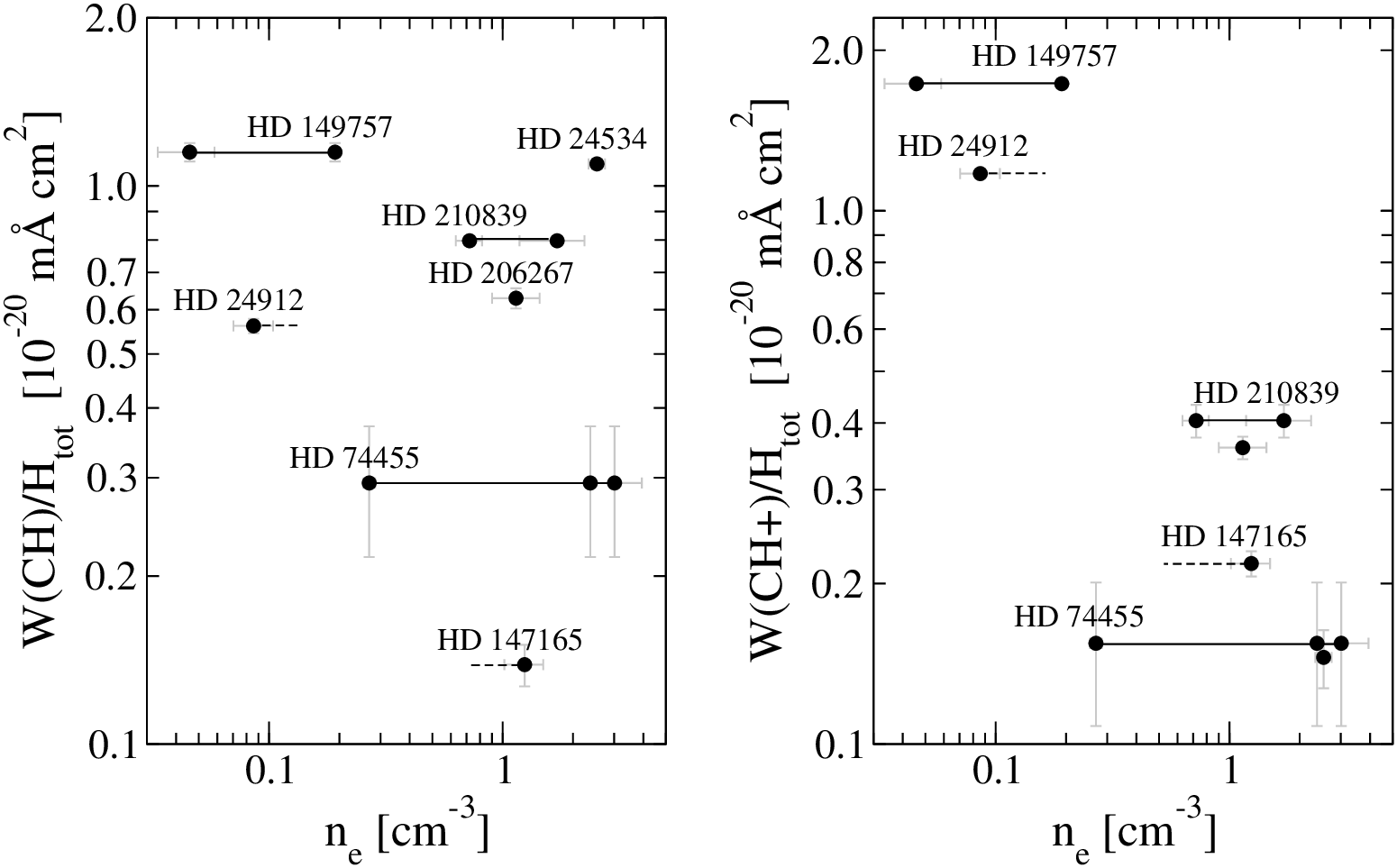}
      \caption{  The equivalent widths of the CH and CH+ spectral lines normalised to the
                 total hydrogen column density are plotted versus electron density
                 ($n_e$). The equivalent width of the CH and CH+ lines includes all                    intervening doppler components.
                 The solid lines connect points from different
                 clouds in the direction of one star. The dashed lines indicates second
                 cloud with undeterminable electron density.
              }
         \label{CH}
   \end{figure*}

\begin{figure*}
   \centering
   \includegraphics[width=\textwidth]{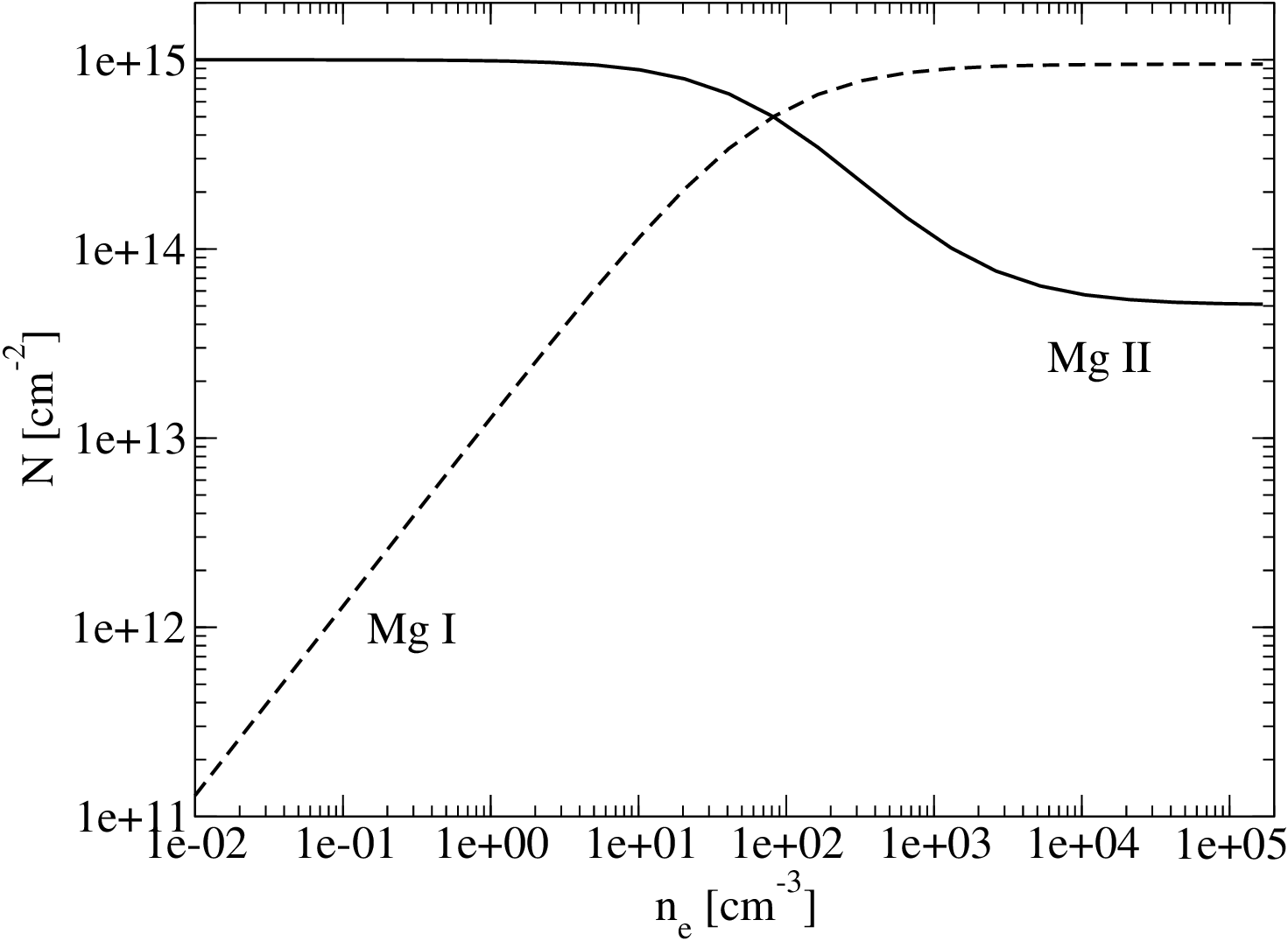}
      \caption{ Theoretical relation between the magnesium column density and 
                the electron density $n_{e}$. 
                Solid line represents Mg II column density, while the dashed line 
                represents Mg I column density. 
              }
         \label{Mg}
         \vspace{1cm}
   \end{figure*}

  Figure \ref{CH} presents the relation between the electron density and equivalent widths of the CH and CH+ molecule normalised to the total hydrogen column density. 
The equivalent widths of CH and CH+ include all doppler components.
The CH abundance does not change between clouds with various electron density. In opposition to CH, the CH+ abundance is lower for clouds with large electron densities (more recombinations). Such behaviour is also illustrated on Fig. \ref{Mg}, where the theoretical relation between the column densities of Mg I and Mg II are presented versus the electron density.

  We have checked that the changes in normalised CH+ equivalent width with $n_e$ are
statistically significant. The points on Fig. \ref{CH} that are in the direction to the same star (connected with a straight line) were replaced by an average $n_e$ value from two
extreme points. The sample of stars was divided in two sets. One with $n_e<0.4$ cm$^{-3}$
and the second one for directions with $n_e>0.4$ cm$^{-3}$. We have calculated the average W(CH)/H$_{tot}$ and W(CH+)/H$_{tot}$ and theirs standard deviations for stars in
both sets. The Student's t-variable was calculated in order to check the agreement between the averages for directions with $n_e<0.4$ cm$^{-3}$ and $n_e>0.4$ cm$^{-3}$. The average of normalised CH for directions with low and high $n_e$ agree with significance level 0.7. The average W(CH+)/H$_{tot}$ for directions with $n_e<0.4$ cm$^{-3}$ and $n_e>0.4$ cm$^{-3}$ differs substantially (significance level 0.009).

\begin{figure*}
   \centering
   \includegraphics[width=\textwidth]{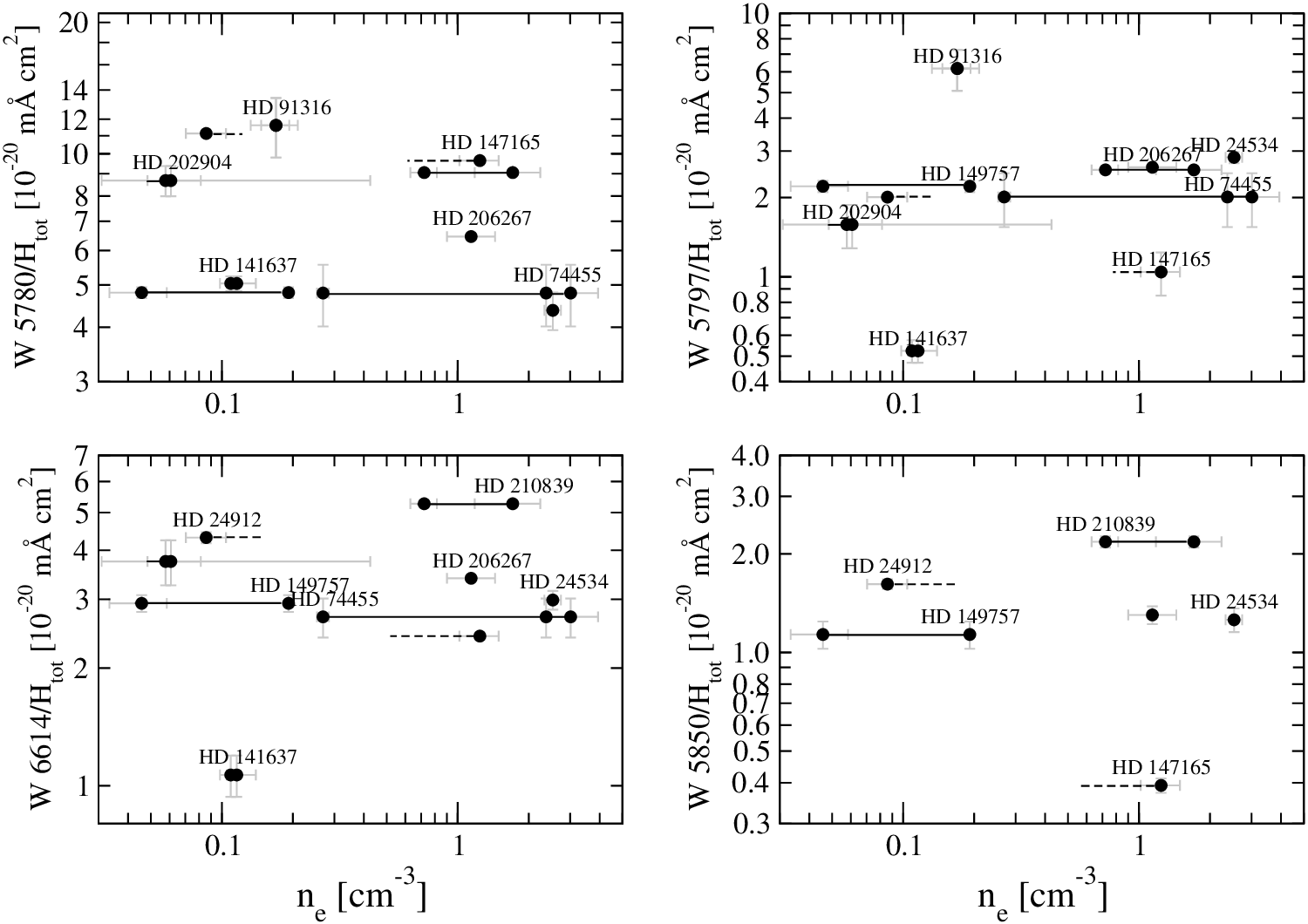}
      \caption{ The equivalent widths of DIBs normalised to the total hydrogen column
                density are plotted versus electron density ($n_e$).
                The solid lines connect points from different clouds in the 
                direction of one star.
                The dashed lines indicates second cloud with undeterminable 
                electron density.       }
         \label{Diby}
   \end{figure*}

\begin{figure*}
   \centering
   \includegraphics[width=\textwidth]{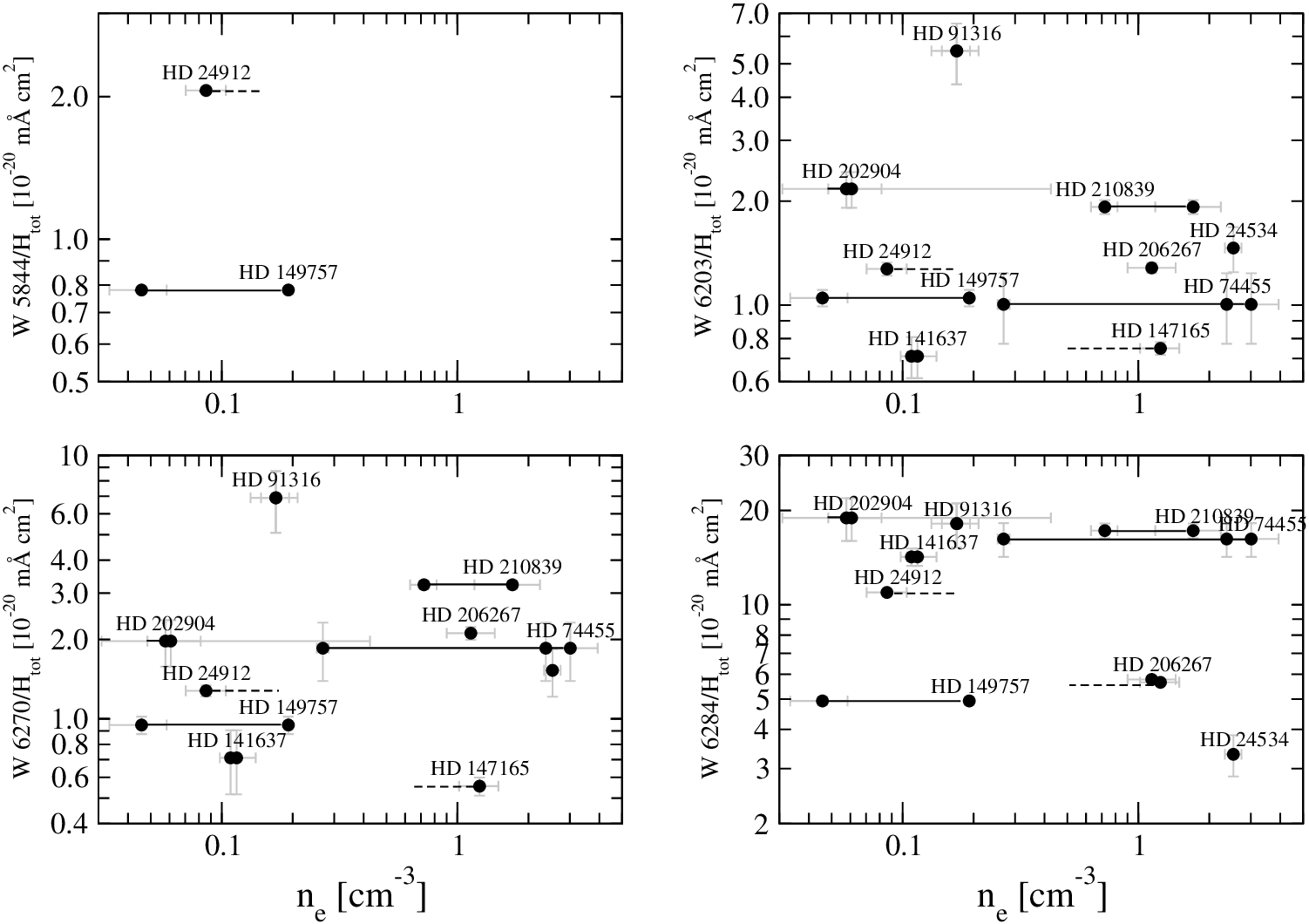}
      \caption{ The equivalent widths of DIBs normalised to the total hydrogen column
                density are plotted versus electron density ($n_e$).
                The solid lines connect points from different clouds in the 
                direction of one star.
                The dashed lines indicates second cloud with undeterminable 
                electron density.       }
         \label{Diby2}
         \vspace{1cm}
   \end{figure*}

\begin{figure*}
   \centering
   \includegraphics[width=\textwidth]{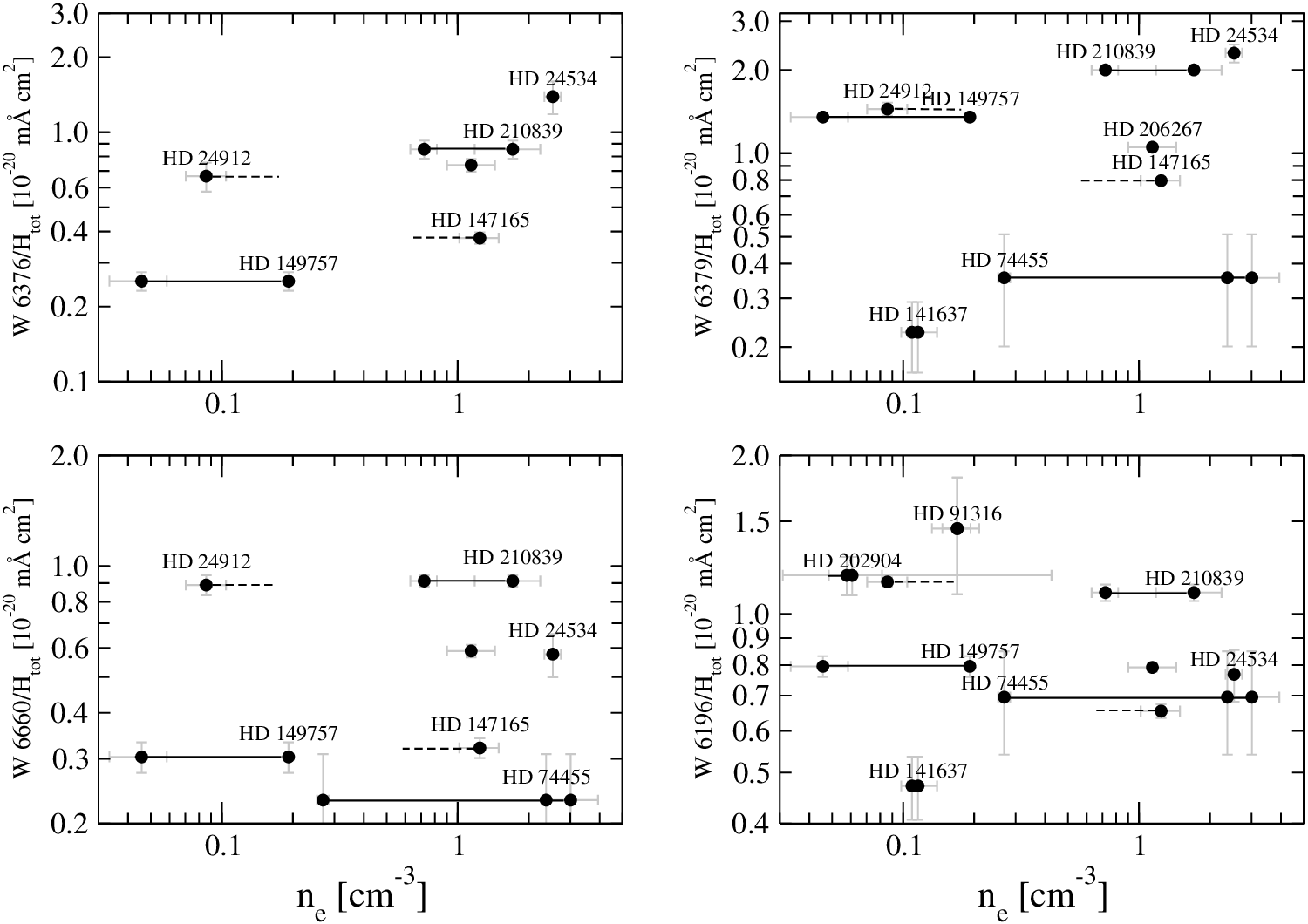}
      \caption{ The equivalent widths of DIBs normalised to the total hydrogen column
                density are plotted versus electron density ($n_e$).
                The solid lines connect points from different clouds in the 
                direction of one star.
                The dashed lines indicates second cloud with undeterminable 
                electron density.      }
      \label{Diby3}
   \end{figure*}

  Figures \ref{Diby}, \ref{Diby2} and \ref{Diby3} presents equivalent widths of DIBs normalised to the total hydrogen column density (N(H$_{tot}$)=N(HI)+2N(H$_2$)) ploted versus electron density ($n_e$ in cm$^{-3}$). One could expect a drop of DIBs equivalent width with $n_e$ as seen on Figure \ref{CH} for CH+. Unfortunately, none of the DIBs bands show a relationship with varying electron density. There are two possible explanations for the lack of relationship between DIBs and $n_e$.
First explanation, that the carriers of the analysed DIBs may be observed only in one ionisation stage. At Figure \ref{Mg} we can see such behaviour for Mg II. In a wide range of observed electron densities ($n_e$=0.009-2.5 cm$^{-3}$) the column density of Mg II does not change by a considerable amount. Such behaviour is also seen for CH on Figure \ref{CH}.

 The second explanation is that DIBs arise in parts of interstellar clouds where we observe only one stage of ionisation of Mg and other elements. The DIBs can arise in dense cores of interstellar clouds, where ionised atoms are hardly observed. DIBs may also arise at outer (ionised) parts of the interstellar clouds, where neutral elements are absent. For both cases we can not calculate the electron density.
The hypothesis that DIBs carriers are formed in outer regions of interstellar clouds was already formulated by \cite{Snow}. They observed that the 4430\AA\  and 5780\AA\ DIBs are
shallower than expected in dense molecular clouds. This result was confirmed by observations of the Taurus dark clouds made by \cite{Adamson}. High UV flux on the clouds surface may be responsible for ionising the DIBs carriers, while the clouds cores are shielded by extinction on dust grains.
  
\section{Conclusions}

 The results can be recapitulated as follows:
   \begin{enumerate}
      \item The electron density in the interstellar clouds was determined for 13 lines of sight.
      \item Linear correlation between the electron density and maximum possible value of electron density from MgI/MgII was found $n_{e}^{MAX}=2.84 \cdot n_{e}$.
      \item The normalised equivalent width of the CH+ line drops with rising electron density as expected from the ionisation equilibrium.
      \item The normalised equivalent widths of the 5780\AA, 5797\AA, 6614\AA, 5850\AA, 5844\AA, 6203\AA, 6270\AA, 6284\AA, 6376\AA, 6379\AA, 6660\AA, 6196\AA\ DIBs do not change with electron density varying in the range $n_e=0.01\div 2.5$ cm$^{-3}$ (diffuse gas).
   \end{enumerate}

\begin{acknowledgements}
   We are very grateful to Jacek Kre\l owski for the equivalent widths of the DIBs and the CH/CH+ lines.
   This publication is based on observations made with the NASA/ESA Hubble Space Telescope, obtained from the data archive at the Space Telescope Science Institute. STScI is operated by the Association of Universities for Research in Astronomy, Inc. under NASA contract NAS 5-26555.

\end{acknowledgements}


\begin{table*}
\begin{tabular}{|l|c|c|c|c|c|c|c|}
\hline \hline
Star & v [$km/s$] & Mg I          & Mg II          &  Si I          & Si II            & C II     & C II*  \\
\hline 
HD 24534  & 18   &$ 7.2\pm0.4e13 $&$ 2.9\pm0.1e15 $&$ 9.3\pm0.6e11 $&$ 1.8\pm0.1e15 $&&\\
HD 24912  & 14   &$ 4.9\pm0.6e12 $&$ 5.5\pm0.4e15 $&$  $&$  $&$  $&$ $\\
HD 74455  & 25   &$ 8.6\pm0.3e12 $&$ 3.1\pm0.1e15 $&$  $&$  $&$  $&$ $\\
HD 74455  &  5   &$ 1.7\pm0.1e12 $&$ 5.5\pm1.1e13 $&$  $&$  $&$  $&$ $\\
HD 74455  & -160 &$ 4.2\pm0.2e11 $&$ 1.7\pm0.1e13 $&$  $&$  $&$  $&$ $\\
HD 91316  & -8   &$ 4.2\pm0.6e12 $&$ 2.4\pm0.1e15 $&$  $&$  $&$  $&$ $\\
HD 91316  & 17   &$ 1.8\pm0.4e12 $&$ 1.0\pm0.1e15 $&$  $&$  $&$  $&$ $\\
HD 141637 & -7   &$ 7.7\pm0.1e12 $&$ 6.8\pm0.2e15 $&$  $&$  $&$  $&$ $\\
HD 141637 & -12  &$ 2.8\pm0.1e12 $&$ 2.3\pm0.4e15 $&$  $&$  $&$  $&$ $\\
HD 147165 & -9   &$ 1.4\pm0.2e13 $&$ 1.1\pm0.1e15 $&$  $&$  $&$  $&$ $\\
HD 149757 & -17  &$ 4.3\pm0.1e12 $&$ 2.1\pm0.1e15 $&$  $&$  $&$  $&$ $\\
HD 149757 & -29  &$ 3.9\pm1.0e11 $&$ 8.1\pm0.2e14 $&$  $&$  $&$  $&$ $\\
HD 160578 & -27  &$ 1.1\pm0.3e11 $&$ 2.3\pm0.3e14 $&$  $&$  $&$ 1.7\pm1.1e16 $&$ 8.0\pm1.3e12 $\\
HD 202904 & -22  &$ 1.3\pm0.1e12 $&$ 3.6\pm1.9e14 $&$  $&$  $&$ 8.4\pm7.4e15 $&$ 2.4\pm0.1e13 $\\
HD 202904 & -13  &$ 1.2\pm0.1e12 $&$ 1.2\pm0.1e15 $&$  $&$  $&$ 1.8\pm0.5e16 $&$ 6.4\pm0.1e13 $\\
HD 203374 & -18  &$ 1.2\pm0.1e14 $&$ 1.2\pm0.2e16 $&$ 1.8\pm0.2e12 $&$ 1.4\pm0.2e16 $&$ $&$ $\\
HD 206267 & -13  &$ 1.7\pm0.1e14 $&$ 1.1\pm0.1e16 $&$ 2.3\pm0.2e12 $&$ 1.0\pm0.1e16 $&$ $&$ $\\
HD 209339 & -15  &$ 1.2\pm0.1e14 $&$ 1.2\pm0.1e16 $&$ 1.2\pm0.3e12 $&$ 9.0\pm2.1e15 $&$  $&$ $\\
HD 210839 & -31  &$ 2.9\pm0.3e13 $&$ 1.9\pm0.1e15 $&$ 4.9\pm1.5e11 $&$ 1.4\pm0.1e15 $&$  $&$ $\\
HD 210839 & -13  &$ 8.5\pm0.1e13 $&$ 1.0\pm0.1e16 $&$ 1.3\pm0.1e12 $&$ 9.0\pm0.3e15 $&$  $&$ $\\
\hline
\end{tabular}
\caption{ Column densities [$cm^{-2}$] derived from the HST spectra. This column densities were used to calculate the electron density ($n_e$).
  }
\end{table*}

\begin{table*}
\begin{tabular}{|c|c|c|c|c|}
  \hline \hline
     & $\Gamma$ & $C(T)$     & $\alpha_{rad}(T)$ & $\alpha_{die}(T)$ \\
     & $[1/s]$  & $[cm^3/s]$ & $[cm^3/s]$        & $[cm^3/s]$ \\
  \hline 
  Mg I/II &$ 8.1\cdot 10^{-11} $&$ 
      \begin{array}{l} 8.9\cdot 10^{-11} \sqrt{T} (1+0.1\cdot T/88700)^{-1} \cdot \\ \cdot exp(-88700/T)  \end{array}   $&$ 
      \begin{array}{l} 1.4\cdot 10^{-13}\cdot \\ \cdot (T/10000)^{-0.855} \end{array} $&$ 
      \begin{array}{l} 4.49\cdot 10^{-4} T^{-3/2} exp(-50100/T)\cdot \\ \cdot (1+0.021\cdot exp(-28100/T)) \end{array} $ \\
  \hline
  Si I/II &$ 3.8\cdot 10^{-9} $&$ 
      \begin{array}{l} 3.92\cdot 10^{-10} \sqrt{T} (1+0.1\cdot T/94600)^{-1}\cdot \\ \cdot exp(-94600/T) \end{array}  $&$ 
      \begin{array}{l} 5.9\cdot 10^{-13}\cdot \\ \cdot (T/10000)^{-0.601} \end{array} $&$ 
      \begin{array}{l} 1.1\cdot 10^{-3} T^{-3/2} exp(-77000/T) \end{array} $ \\
  \hline
   Ca I/II &$ 3.8\cdot 10^{-10} $&$ 
      \begin{array}{l} 2.09\cdot 10^{-10} \sqrt{T} (1+0.1\cdot T/70900)^{-1}\cdot \\ \cdot exp(-70900/T) \end{array}  $&$ 
      \begin{array}{l} 1.12\cdot 10^{-13}\cdot \\ \cdot (T/10000)^{-0.9} \end{array} $&$ 
      \begin{array}{l} 3.28\cdot 10^{-4} T^{-3/2} exp(-34600/T) \\ \cdot (1+0.0907\cdot exp(-16400/T)) \end{array} $ \\
  \hline
   C II/II* & -- &$ 
      \begin{array}{l} 8.63\cdot 10^{-6} (2\sqrt{T})^{-1} \Omega_{12}(T) \cdot \\ \cdot exp(-\frac{1.31\cdot 10^{-14} erg}{kT}) \end{array}  $&$ 
      2.29\cdot 10^{-6} $&
      -- \\
  \hline
\end{tabular}
\caption{ Parameters used in the ionisation equilibrium equation. The ionisation rate $\Gamma$ was adopted from the WJ2 model of \cite{Boer}. The recombinations rates $\alpha_{rad}$ and $\alpha_{die}$ as well as the collisional ionisation rate $C$ for Mg,
 Si and Ca was adopted from \cite{Shull}. The $\alpha_{rad}(CII)$ comes from \cite{NS}.
 The collision strength $\Omega_{12}(T)$ was fited by an 8-order polynomial to the data given by \cite{Hayes}. }
 \vspace{1cm}
\end{table*}

\begin{table*}
\begin{tabular}{|l|r|r|r|rr|rr|r|r|r|}
\hline \hline
Star & v & n$_e$          & n$_{e}^{MAX}$  &  HI                        & ref & H$_2$                 & ref & CH$^+$  & CH  \\
     &[$km/s$]        & [$cm^{-3}$] & [$cm^{-3}$]  & \multicolumn{2}{|l|}{[$log N(HI)$]} &         \multicolumn{2}{|l|}{[$log N(H_2)$]}           & [m\AA]   & [m\AA]   \\
\hline 
HD 24534  & 18  & $2.5   ^{+0.2}  _{-0.2}$  & $6.7 ^{+0.6} _{-0.6}$ & 20.73 & 3 & 20.92 & 1 & $3.2 \pm 0.4$ & $24.1 \pm 0.5$  \\
HD 160578 & -27 & $0.009 ^{+0.02}_{-0.004}$ & $0.14 ^{+0.06} _{-0.05}$ & 20.19 & 3 & & & &  \\
HD 202904 & -22 & $0.06  ^{+0.38} _{-0.03}$  & $0.9 ^{+1.3} _{-0.4}$ & 20.68 & 4 & 19.15 & 4 & & \\
HD 202904 & -13 & $0.06  ^{+0.02} _{-0.01}$  & $0.26 ^{+0.04} _{-0.04}$ & & & & & &  \\
HD 203374 & -18 & $0.66  ^{+0.22} _{-0.16}$  & $2.7 ^{+0.5} _{-0.4}$ & & & & & &   \\
HD 206267 & -13 & $1.1   ^{+0.3}  _{-0.2}$  & $4.3 ^{+0.2} _{-0.2}$ & 21.30 & 5 & 20.86 & 1 & $11.3 \pm 0.8$ & $21.7 \pm 0.9$  \\
HD 209339 & -15 & $0.7   ^{+0.4}  _{-0.3}$   & $2.7 ^{+0.2} _{-0.2}$ & & & & & &  \\
HD 210839 & -31 & $1.7   ^{+0.5}  _{-0.5}$   & $4.1 ^{+0.6} _{-0.5}$ & 21.15 & 3 & 20.84 & 1 & $11.3 \pm 0.8$ & $22.3 \pm 0.4$  \\
HD 210839 & -13 & $0.72  ^{+0.1} _{-0.09}$  & $2.24 ^{+0.08} _{-0.07}$ & & & & & & \\
\hline
\multicolumn{10}{|l|}{Directions with the electron density calculated with the formula $n_{e}=n_{e}^{MAX}/2.84$ from the Mg column densities.}  \\
\hline
HD 24912  & 14  & $0.09  ^{+0.02} _{-0.02}$  & $0.24 ^{+0.05} _{-0.04}$ & 21.05 & 3 & 20.53 & 2 & $21.13 \pm 0.19$ & $10.1 \pm 0.3$  \\
HD 74455  & 25   & $0.27 ^{+0.02} _{-0.02}$ & $0.76 ^{+0.05} _{-0.04}$ & 20.73 & 3 & 19.74 & 6 & $1.0 \pm 0.3$ & $1.9 \pm 0.5$ \\
HD 74455  &  5   & $3.0 ^{+0.9} _{-0.6}$ & $8.6 ^{+2.6} _{-1.7}$ & & & & & & \\
HD 74455  & -160 & $2.4 ^{+0.09} _{-0.08}$ & $6.7  ^{+0.3} _{-0.2}$ & & & & & & \\
HD 91316  & -8 & $0.17 ^{+0.02} _{-0.02}$ & $0.48  ^{+0.07} _{-0.07}$ & 20.44 & 3 & 15.61 & 7 & & \\
HD 91316  & 17 & $0.17 ^{+0.04} _{-0.04}$ & $0.5  ^{+0.1} _{-0.1}$ & & & & & & \\
HD 141637 & -7 & $0.109 ^{+0.005} _{-0.004}$ & $0.31  ^{+0.01} _{-0.01}$ & 21.18 & 3 & 19.23 & 4 & & \\
HD 141637 & -12 & $0.12 ^{+0.02} _{-0.02}$ & $0.33  ^{+0.07} _{-0.05}$ & & & & & & \\
HD 147165 & -9  & $1.2   ^{+0.3}  _{-0.2}$   & $3.5 ^{+0.7} _{-0.6}$ & 21.38 & 3 & 19.79 & 2 & 4.5 & 2.9  \\
HD 149757 & -17 & $0.192 ^{+0.005}_{-0.006}$ & $0.54 ^{+0.02} _{-0.02}$ & 20.69 & 3 & 20.65 & 2 & 22.4 & 18.0  \\
HD 149757 & -29 & $0.046 ^{+0.013}_{-0.012}$ & $0.13 ^{+0.04} _{-0.04}$ & & & & & & \\
\hline
\end{tabular}
\caption{The electron densities derived for target stars. The equivalent widths of the CH$^+$ and CH lines were adopted from \cite{Kre} and Kre\l owski (priv. comm.).
These equivalent widths include absorption for all doppler components.
 References: (1)-\cite{Rach} ; (2)-\cite{Sav} ; (3)-\cite{Dip} ; (4)-\cite{Jen} ; (5)-\cite{Lacour} ;
 (6) - N(H$_2$) calculated from the formula : N(H$_2)=2.9e19 \cdot$ W(CH);
 (7) -\cite{Boh} ; 
. \label{tabela}
  }
\end{table*}

\begin{table*}
\scriptsize
\begin{tabular}{lrrrrrrrrrrrr}
\hline \hline
Star      & 5797           & 5780             & 5850           & 5844 & 6196           & 6203           & 6270           & 6284            & 6376           & 6379           & 6614            & 6660  \\
\hline
HD 24534  & 62.5 $\pm$ 3.1 & 96.2  $\pm$ 9.5  & 27.7 $\pm$ 2.3 &      & 16.9 $\pm$ 1.9 & 32.2 $\pm$ 4.8 & 33.6 $\pm$ 6.9 & 73.3  $\pm$ 11  & 30.6 $\pm$ 4.6 & 50.6 $\pm$ 3.8 & 65.7  $\pm$ 3.6 & 12.7 $\pm$ 1.7 \\
HD 202904 & 8    $\pm$ 1.5 & 44    $\pm$ 3.5  &                &      & 6    $\pm$ 0.5 & 11   $\pm$ 1.3 & 10   $\pm$ 2   & 96    $\pm$ 15  &                &                & 19    $\pm$ 2.5 &           \\
HD 206267 & 89.8 $\pm$ 1.6 & 222.7 $\pm$ 3.6  & 44.9 $\pm$ 2.8 &      & 27.3 $\pm$ 0.5 & 44.2 $\pm$ 1.5 & 72.9 $\pm$ 4.1 & 199.2 $\pm$ 4.6 & 25.5 $\pm$ 1.5 & 36.3 $\pm$ 0.9 & 117   $\pm$ 1.1 & 20.3 $\pm$ 0.8 \\
HD 210839 & 71.2 $\pm$ 0.9 & 253.1 $\pm$ 2.7  & 60.9 $\pm$ 2.4 &      & 30.7 $\pm$ 1.1 & 53.8 $\pm$ 2.5 & 90.2 $\pm$ 3.3 & 482   $\pm$ 26  & 23.9 $\pm$ 2   & 55.9 $\pm$ 1.3 & 147.2 $\pm$ 2.5 & 25.5 $\pm$ 0.9 \\
HD 24912  & 36.1 $\pm$ 0.6 & 200.3 $\pm$ 2.4  & 29.1 $\pm$ 1   & 37.1 & 20.7 $\pm$ 0.4 & 22.9 $\pm$ 1   & 23   $\pm$ 1.1 & 197   $\pm$ 3.5 & 12   $\pm$ 1.6 & 26   $\pm$ 1.4 & 77.6  $\pm$ 2.5 & 16   $\pm$ 1 \\
HD 74455  & 13   $\pm$ 3   & 31    $\pm$ 5    &                &      & 4.5  $\pm$ 1   & 6.5  $\pm$ 1.5 & 12   $\pm$ 3   & 105   $\pm$ 13  &                & 2.3  $\pm$ 1   & 17.5  $\pm$ 2   & 1.5  $\pm$ 0.5 \\
HD 91316  & 17   $\pm$ 3   & 32    $\pm$ 5    &                &      & 4    $\pm$ 1   & 15   $\pm$ 3   & 19   $\pm$ 5   & 50    $\pm$ 8   &                &                &                 &          \\
HD 141637 & 8.1  $\pm$ 0.8 & 78    $\pm$ 3    &                &      & 7.3  $\pm$ 1   & 11   $\pm$ 1.5 & 11   $\pm$ 3   & 220   $\pm$ 14  &                & 3.5  $\pm$ 1   & 16.5  $\pm$ 2   &          \\
HD 147165 & 26.3 $\pm$ 4.9 & 243.3 $\pm$ 3.1  & 9.9  $\pm$ 0.5 &      & 16.5 $\pm$ 0.5 & 18.9 $\pm$ 0.8 & 14   $\pm$ 1.1 & 142.6 $\pm$ 2.1 & 9.5  $\pm$ 0.5 & 20.1 $\pm$ 0.3 & 60.9  $\pm$ 1.2 & 8.1  $\pm$ 0.5 \\
HD 147933 & 50.8 $\pm$ 2.4 & 208   $\pm$ 12.5 & 28.1 $\pm$ 0.8 & 20.3 & 16.6 $\pm$ 0.7 & 22.8 $\pm$ 1.3 & 20   $\pm$ 1.3 & 176.4 $\pm$ 2.8 & 11.6 $\pm$ 0.8 & 25.9 $\pm$ 1   & 64.6  $\pm$ 1.7 & 11.8 $\pm$ 0.9 \\
HD 149757 & 30.5 $\pm$ 1.5 & 66.4  $\pm$ 1.9  & 15.7 $\pm$ 1.5 & 10.8 & 11   $\pm$ 0.5 & 14.5 $\pm$ 0.8 & 13.1 $\pm$ 1   & 68.2  $\pm$ 2   & 3.5  $\pm$ 0.3 & 18.7 $\pm$ 0.5 & 40.5  $\pm$ 2   & 4.2  $\pm$ 0.4 \\
\hline
\end{tabular}
\caption{ \small Equivalent widths of the diffuse interstellar bands in m\AA \ (courtesy of Jacek Kre\l owski).}
\end{table*}


\end{document}